\documentclass[a4paper]{article}
\usepackage{aqis}

\begin{document}

\title{Role of Hypothesis Testing in Quantum Information}

\author{
Masahito Hayashi \affiliation{1}
   \affiliation{2}
  \email{masahito@math.nagoya-u.ac.jp}
  }
  \address{1}{
Graduate School of Mathematics, Nagoya University, Nagoya, 464-8602, Japan}
\address{2}{
Centre for Quantum Technologies, National University of Singapore, 3 Science Drive 2, 117542, Singapore}

\abstract{Recently, it is well recognized that hypothesis testing has deep relations with other topics in quantum information theory as well as in classical information theory.
These relations enable us to derive precise evaluation in the finite-length setting.
However, such usefulness of hypothesis testing is not limited to information theoretical topics. For example, it can be used for verification of 
entangled state and quantum computer as well as 
guaranteeing the security of keys generated via quantum key distribution.
In this talk, we overview these kinds of applications of hypothesis testing.}


\maketitle


\section{Quantum information theory and binary hypothesis testing}
In information theory community
it is well known that 
many information theoretical tasks can be analyzed by using the terminology of the binary asymmetric hypothesis testing.
While there are many studies to focus on this relation,
the first series study with this direction is the method of information spectrum, which was initiated by Han and Verd\'{u} \cite{Han1}\cite{resolvability}\cite{Verdu-Han}, in which
we convert the optimization problems in various information tasks
into the binary asymmetric hypothesis testing, and
the asymptotic behavior of the likelihood ratio plays a key role.
This correspondence is valid without any assumption for the information source and/or the information channel,
i.e., we do not need the independent and identical distributed condition nor Markovian condition.
Due to the generality of the method of information spectrum, 
Nagaoka \cite{Nagaoka} considered to employ this method for quantum information theory.
As a result, he found a remarkable relation between the classical-quantum channel coding and the quantum binary hypothesis testing, in which the correctly decoding probability is upper bounded by 
the performance of the corresponding quantum binary hypothesis testing in a canonical way.
Later, Polyanskiy, Poor and Verd\'{u} \cite{Pol} showed the same inequality only with the classical channel coding, 
which is called meta-converse theorem, nowadays.
Nagaoka \cite{Nagaoka} also pointed out the notable relation between the quantum binary hypothesis testing 
and the R\'{e}nyi relative entropy.
These his results were presented 
in the first conference of ERATO Workshop on Quantum Information Science, which is the forerunner of AQIS conference series \cite{Nagaoka}.

Based on this study, the author jointly with Nagaoka proved 
another remarkable relation between the classical-quantum channel coding and the quantum binary hypothesis testing \cite{HN,Ha09-8}.
That is, they showed that the decoding error probability is upper bounded by 
the error probability of the corresponding quantum binary hypothesis testing, which is chosen slightly differently from the meta converse.
Based on this method, the author derived the lower bound of the error exponent in the classical-quantum channel \cite{exponent-cq}.
Wang and Renner \cite{Wang} reformulated this result by introducing the hypothesis testing entropy. 
Later, Polyanskiy, Poor and Verd\'{u} showed the same inequality only with the classical channel coding, which is called the dependence test (DT) bound \cite{Pol}.
These two remarkable relations lead the breakthrough of the second order analysis of channel coding \cite{Second,Pol}. 

Also, quantum data compression 
can be treated via the quantum binary hypothesis testing \cite{source-co,Nag-Hay}.
Since reduction to a quantum analogue of likelihood ratio test, i.e., the quantum binary hypothesis testing 
is a very powerful method \cite{Springer,Ha09-8},
the following topics can be treated in this direction;
quantum wiretap channel \cite{wire}, 
universal (compound) channel coding \cite{uni,compound},
entanglement concentration \cite{concent}, 
entanglement dilution \cite[Section 8.6]{Springer}, 
classical data compression with quantum side information \cite{Tomamichel},
quantum Slepian-Wolf problem \cite{Slepian-Wolf},
classical random number generation with quantum side information \cite{Tomamichel}, 
quantum state redistribution \cite{redis}, and
entanglement assisted communication over (quantum-quantum) point to point quantum channel, Gel'fand- Pinsker quantum channel, and quantum broadcast channel \cite{ent-a}.


\section{Verification of bipartite entangled state}
Application of quantum hypothesis testing is not limited to
the above type of theoretical aspects.
Quantum hypothesis testing can be applied to more practical topics.
One is verification of a bipartite entangled state.
When an entangled state is generated experimentally,
to use the generated bipartite entangled state,
we need to verify whether the generated state is truly the intent bipartite entangled state.
In the conventional qubit system,
our verification is written as a binary POVM on the bipartite system.
In this case, the direction of the error cannot be expected,
it is suitable that the performance of this testing does not depend on the direction.
That is, the POVM of the testing is preferred to be invariant for the group action preserving the entangled state.
Such a testing method is formulated by using 
the irreducible decomposition of the group representation theory \cite{Tsuda,LOCC}.

However, in a usual optical device, like, spontaneous parametric down-conversion (SPDC),
 a binary POVM is often constructed by a filter and a detector.
That is, when the filter is passed, we have detection.
Otherwise, we have no detection.
In this situation, it is impossible to distinguish the following two cases, both of which correspond to no detection.
One is the event that the photon pair is not generated so that it is not detected.
The other is the event that the photon pair is generated, but the filter is not passed so that it is not detected.
Then, the performance of the following two cases are not the same.
We have the detection when the generated state is close to the intent entangled state.
We have the detection when the generated state is far from the intent entangled state.
Surprisingly, when the photon generation rate is known, the latter has better performance for this testing \cite{Tomita}, whose experimental demonstration was also done \cite{Jiang}.

\section{Quantum key distribution}
Another important application of hypothesis testing is its application to 
quantum key distribution, which is a method to share secure keys via quantum communications
and classical communications \cite{BB84}.
Its security is trivial when the quantum communication channel has no noise.
However, a real quantum communication channel has a certain amount of noise.
When the amount is less than a threshold,
we can generate secure keys by combining the error correction and the privacy amplification \cite{SP}.
Quantum key distribution focuses on the bit basis and the phase basis.
The error of the bit basis expresses the sacrifice rate in the error correction
and the error of the phase basis expresses the sacrifice rate in the privacy amplification \cite{H-QKD}.
Later, a similar observation was also done via smooth entropy \cite{To}.
While we randomly choose check bits in quantum key distribution, 
its purpose is verification of the error rates of both bases.

However, in a realistic quantum key distribution, we often employ weak coherent pulses,
which generates multi-photon state with some probability.
In this situation, only a part of generated photons arrive at the receiver side.
When a multi-photon state is generated, the eavesdropper, in principle,
can obtain the transmitted information.
Therefore, the required sacrifice rate in the privacy amplification 
is determined by 
the rate of pulses generated as multi-photon among the pluses arrived at the receiver side
and the error rate of phase basis among pulses generated as single-photon and arrived at the receiver side \cite{aa}. 
That is, when we employ weak coherent pulses,
we need to know these two ratios as well as the bit error ratio of the received pulses.
For this purpose, we need to guarantee that 
these two rates are not greater than certain values.
In this verification process, we randomly chooses several values of intensity of pulses \cite{HW}.
Then, we obtain the detection rate and the error rates of the phase basis depending on the intensity.
Using this data, we apply the method of hypothesis testing or 
the interval estimation, which employs the percent point \cite{Nakayama}.
Hence, we can verify these two ratios with certain intervals.

\section{Verification of quantum computer}
Hypothesis testing can be applied to the verification of quantum computer.
In the conventional circuit model, it is quite difficult to predict the outcome of the circuit because the aim of the computation is to know the outcome.
As an alternative model of quantum computer, 
measurement-based quantum computer (MBQC) is known \cite{RB}, and 
is composed of a limited number of local measurements and a graph state, which is an entangled state of  large size.
Since the components of MBQC have known forms,
its verification can be done by verifying these components.
Therefore, when we can trust these local measurement devices,
we can verify our computation outcome under the MBQC model by verifying the graph state.
However, since available measurements are restricted to a limited number of local measurements,
we need to realize the verification only with this limited class of measurements.
Fortunately, in the graph state, 
the outcome of the $Z$ basis predicts the behavior of the connected site.
In the case of two-colorable graph, 
we can deterministically predict the outcome of the $X$ basis on sites of one color 
from the $Z$ basis outcome on sites of the other color.
Using this property, we can verify whether the generated state is the intent graph state \cite{Morimae}.
Since the above prediction is deterministic, this verification is can be done very efficiently.
That is, the required number of sampling does not depend on the size of the graph state.
Since this test checks whether the state belongs to the stabilizer defined by 
the pair of the $X$ basis measurement and the $Z$ basis measurement,
it is called the stabilizer test.

Further, this method can be extended to the case when the measurement devices has noises
and the generated graph state has noise \cite{Fujii}.
In this case, we need to attach the fault-tolerant MBQC, which is often based on a topological surface code.
When the noises of measurement devices are independent, they can be theoretically converted 
to the noises in the generated graph state.
Once we fix the scheme of the fault-tolerant MBQC, we can define the set of correctable errors.
When the noise belongs to the set of correctable error,
the fault-tolerant MBQC works properly, i.e., the computation outcome is the correct value.
Hence, it is sufficient to verify whether the error belongs to the the set.
Since this test is also deterministic, 
the required number of sampling still does not depend on the size of the graph state.

Furthermore, we can make this kind of test even when the measurement device cannot be trusted.
This kind of test is called self-testing, and 
the currently proposed method works with the noiseless case.
In this setting, the testing of a graph state can be reduced to the testing of the Bell state in a canonical way.
McKague et al \cite{McKague} proposed the self-testing of the Bell state only with the CHSH test.
However, the recently proposed method \cite{Hajdusek} combines the CHSH test and the stabilizer test so that
the performance is much improved.
This self-testing of the Bell state yields a self-testing of the graph state. 
When it applied to the verification of MBQC, 
the obtained scaling is much better than previously obtained verification methods \cite{RUV,McKague2}, 
and is the same as that of the paper \cite{Fitzsimons}, which employs a different method.

\section*{Acknowledgements}
The author is grateful to all the collaborators of these presented results.
Also, he is thankful for all the financial supports of these presented results.


\begin{thebibliography}{9}
\bibitem{ent-a}
A. Anshu, R. Jain, and N. A. Warsi,
arXiv:1702.01940 2017.

\bibitem{redis}
A. Anshu, R. Jain, and N. A. Warsi,
arXiv:1702.02396 2017.

\bibitem{Slepian-Wolf}
A. Anshu, R. Jain, and N. A. Warsi,
arXiv:1703.09961 2017.

\bibitem{compound}
A. Anshu, R. Jain, and N. A. Warsi,
arXiv:1706.08286 2017.

\bibitem{BB84}
C. H. Bennett and G. Brassard,
Quantum cryptography: public key distribution and coin tossing,
{\em Proc. IEEE International Conference on Computers, Systems and
Signal Processing},
(Bangalore, India), pp.~175--179, 1984.

\bibitem{Fujii}
K. Fujii and {M. Hayashi},
Verifiable fault tolerance in measurement-based quantum computation,
{\em Phys. Rev. A, Rapid Communication}, Accepted; arXiv:1610.05216.

\bibitem{Fitzsimons}
M. Hajdu\v{s}ek, C. A. P\'{e}rez-Delgado, and J. F. Fitzsimons,
arXiv:1502.02563, 2015.

\bibitem{Han1}
T.-S. Han, 
{\em Information-Spectrum Methods in Information Theory},
Springer, Berlin, 2003. 
(Original Japanese version: Baifukan, 1998)

\bibitem{resolvability}
T. S. Han and S. Verd\'{u}, 
{\em IEEE Trans. Inf. Theory},
{\bf 39} 752--772, 1993.

\bibitem{source-co}	 
{M. Hayashi}, 
{\em Phys. Rev. A}, {\bf 66} 032321, 2002.

\bibitem{Ha09-8}	 
{M. Hayashi}, 
Hypothesis testing approach to quantum information theory, 
{\em Proceeding of COE Symposium on Quantum Information Theory}, Kyoto, Japan, September 2--3, 2003.

\bibitem{concent}	 
{M. Hayashi}, 
{\em IEEE Trans. Inf. Theory},
{\bf 52} 1904--1921, 2006.

\bibitem{H-QKD}
M. Hayashi, 
{\em Phys. Rev. A}, {\bf 74}, 022307, 2006.

\bibitem{aa}
M. Hayashi, 
{\em Phys. Rev. A}, {\bf 76} 012329, 2007.

\bibitem{exponent-cq}	 
{M. Hayashi}, 
{\em Phys.  Rev. A}, {\bf 76}, 062301, 2007.

\bibitem{Second}	 
{M. Hayashi}, 
{\em IEEE Trans. Inf. Theory}, {\bf 55}  4947--4966, 2009. 

\bibitem{LOCC}	 
{M. Hayashi}, 
{\em New J.  Phys.}, {\bf 11}  043028, 2009.

\bibitem{uni}
M. Hayashi,
{\em Comm. Math. Phys.}, {\bf 289} 1087-1098, 2009.

\bibitem{wire}
{M. Hayashi}, 
{\em IEEE Trans. Inf. Theory}, {\bf 61} 5595--5622, 2015. 

\bibitem{Springer}
{M. Hayashi},
{\em Quantum Information Theory: Mathematical Foundation}, 
{\em Graduate Texts in Physics}, Springer, 2017.
(First edition: Springer, 2006).

\bibitem{Hajdusek}
M. Hayashi and M. Hajdu\v{s}ek,
arXiv:1603.02195, 2016.

\bibitem{Tsuda}
{M. Hayashi}, K. Matsumoto, and Y. Tsuda, 
{\em J. Phys. A: Math. Gen.} {\bf 39} 14427 -- 14446, 2006.

\bibitem{Morimae}
{M. Hayashi} and T. Morimae, 
{\em Phys. Rev. Lett.,} {\bf 115} 220502, 2015. 

\bibitem{HN}	 
{M. Hayashi} and H. Nagaoka, 
{\em IEEE Trans. Inf. Theory}, {\bf 49}  1753 -- 1768, 2003.

\bibitem{Nakayama}
{M. Hayashi} and R. Nakayama, 
{\em New. J. Phys.}, {\bf 16} 063009, 2014.



\bibitem{Jiang}
{M. Hayashi}, B.-S. Shi, A. Tomita, K. Matsumoto, Y. Tsuda, and Y.-K. Jiang, 
{\em Phys. Rev. A}, {\bf 74} 062321, 2006.

\bibitem{Tomita}	
{M. Hayashi}, A. Tomita, and K. Matsumoto, 
{\em New J.  Phys.}, {\bf 10} 043029, 2008.

\bibitem{HW}
W.-Y. Hwang, 
{\em Phys. Rev. Lett.} {\bf 91} 057901, 2003.

\bibitem{McKague2}
M. McKague, {\em Theory of Computing} {\bf 12} 1, 2016.

\bibitem{McKague}
M. McKague, T. H. Yang, and V. Scarani, 
{\em J. Phys. A: Math. Theor.}, {\bf 45} 455304, 2012.

\bibitem{Nagaoka}	 
H. Nagaoka, Strong converse theorems in quantum information theory, 
In {\em Proceedings of ERATO Workshop on Quantum Information Science 2001}, Univ. Tokyo, Tokyo, Japan, September 6-8, 2001, pp. 33.

\bibitem{Nag-Hay}
H. Nagaoka and M. Hayashi,
{\em IEEE Trans. Inf. Theory},
{\bf 53} 534--549, 2007.

\bibitem{Pol}
Y. Polyanskiy, H.V. Poor, and S. Verd\'{u}, 
{\em IEEE Trans. Inf. Theory},
{\bf 56} 2307 -- 2359, 2010.

\bibitem{RB}
R. Raussendorf and H. J. Briegel, 
{\em Phys. Rev. Lett.} {\bf 86} 5188 (2001).

\bibitem{RUV}
B. W. Reichardt, F. Unger, and U. Vazirani,
{\em Nature} {\bf 496} 456, 2013.

\bibitem{SP}
P. W. Shor and  J. Preskill,
{\em Phys. Rev. Lett.}, {\bf 85} 441--444, 2000.

\bibitem{Tomamichel}
M. Tomamichel and {M. Hayashi}, 
{\em IEEE Trans. Inf. Theory},
{\bf 59}  7693--7710, 2013.

\bibitem{To}
M. Tomamichel, C. C. W. Lim, N. Gisin, and R. Renner,
{\em Nat. Com.} {\bf 3} 634, 2012.

\bibitem{Verdu-Han}
S. Verd\'{u} and T. S.\ Han,
{\em IEEE Trans. Inf. Theory},
{\bf 40} 1147--1157, 1994.

\bibitem{Wang}	 
L Wang and R Renner,
{\em Phys. Rev. Lett.}, {\bf 108} 200501, 2012.

\end{thebibliography}
\end{document}